\documentclass[aps,prb,twocolumn,showpacs,preprintnumbers,amsmath,amssymb,superscriptaddress,bibnotes]{revtex4}

\usepackage{graphicx}
\usepackage{dcolumn}
\usepackage{bm}

\begin{document}


\title{Superconducting order parameter in nonmagnetic borocarbides RNi$_{2}$B$_{2}$C (R=Y, Lu) probed by point-contact Andreev reflection spectroscopy}

\author{X. Lu}\altaffiliation[Present address: ]{Los Alamos National Laboratory, Los Alamos, NM, USA.}
\author{W. K. Park}\email[Corresponding author: ]{wkpark@illinois.edu}\affiliation{Department of Physics and the Frederick Seitz Material Research Laboratory, University of Illinois at Urbana-Champaign, Urbana, Illinois 61801, USA}
\author{S. Yeo}\altaffiliation[Present address: ]{Korea Atomic Energy Research Institute, Daejeon 305-353, Republic of Korea.}
\author{K.-H. Oh}
\author{S.-I. Lee}\altaffiliation{Deceased.}\affiliation{National Creative Research Initiative Center for Superconductivity and Department of Physics, Pohang University of Science and Technology, Pohang 790-784, Republic of Korea}
\author{S. L. Bud\textquoteright ko}
\author{P. C. Canfield}\affiliation{Ames Laboratory and Department of Physics and Astronomy, Iowa State University, Ames, Iowa 50011, USA}
\author{L. H. Greene}\affiliation{Department of Physics and the Frederick Seitz Material Research Laboratory, University of Illinois at Urbana-Champaign, Urbana, Illinois 61801, USA}

\date{\today}

\begin{abstract}
We report on the measurements of the superconducting order parameter in the nonmagnetic borocarbides LuNi$_2$B$_2$C and YNi$_2$B$_2$C. Andreev conductance spectra are obtained from nanoscale metallic junctions on single crystal surfaces prepared along three major crystallographic orientations: [001], [110], and [100]. The gap values extracted by the single-gap Blonder-Tinkham-Klapwijk model follow the BCS predictions as a function of temperature and magnetic field and exhibit a small anisotropy. These observations are robust and reproducible among all the measurements on two different sets of LuNi$_2$B$_2$C crystals and one set of YNi$_2$B$_2$C crystals. We suggest possible explanations for the small gap anisotropy based on the recent detailed gap measurements by angle-resolved photoemission spectroscopy and the tunneling cone effect. Our results provide a consistent picture of the superconducting gap structure in these materials, addressing the controversy particularly in the reported results of point-contact Andreev reflection spectroscopy.
\end{abstract}

                            
\maketitle

\section{Introduction}

Discovered more than fifteen years ago,\cite{cava94,nagarajan94} the family of quaternary intermetallic compounds, RNi$_2$B$_2$C (R = rare earth elements), have attracted great attention.\cite{KHMuller-2001Review, Thalmeier-2003, canfield98} They provide a unique opportunity for the investigation of the interplay between superconductivity and magnetism in a homologous series of compounds. They can be classified into four subgroups: magnetic members (R = Pr, Nd, Gd, Tb), superconductors (R = Sc, Y, Lu), coexisting materials (R = Dy, Ho, Er, Tm), and heavy fermions (R = Yb).\cite{KHMuller-2001Review, Thalmeier-2003, canfield98} The non-magnetic superconducting borocarbides, LuNi$_2$B$_2$C and YNi$_2$B$_2$C, a pristine model to study the superconductivity in this family, have the highest superconducting transition temperature: $T_\textrm{c} \sim$ 16.5 K and 15.5 K, respectively. Despite intensive studies over the past decade, there is no consensus on their detailed gap structures. How magnetism and superconductivity coexist or compete in the coexisting members remains a topic of considerable research interest since a systematic evolution of both phase transitions is observed over a comparable temperature range ($1 - 20$ K). On a broader context, the interplay between superconductivity and magnetism has become a central theme in the studies of novel and unconventional superconductors including heavy fermions, cuprates, and the recently discovered iron-based superconductors. The magnetic and superconducting orders in RNi$_2$B$_2$C have different origins, arising from the localized 4f electrons on R$^{3+}$ ions coupled via the Ruderman-Kittel-Kasuya-Yosida (RKKY) interaction,\cite{KHMuller-2001Review, canfield98} whereas the superconducting condensation is associated with the itinerant electron bands residing on the Ni$_2$B$_2$ layers.\cite{Thalmeier-2003}

Complex Fermi surface (FS) structures with multiple sheets have been predicted from band structure calculations\cite{Drechsler-20001-364, Yamauchi-2004-412} and confirmed by several experimental techniques.\cite{Dugdale-1999-83, SBDugdale-09SuperSciTech, starowicz-1, BBergk-2008PRLdHvA} There are three bands crossing the Fermi level, all having electron-like character.\cite{Yamauchi-2004-412, BBergk-2008PRLdHvA} While the small ellipsoidal FS from the 19th band is known to contribute very little to the electronic density of states, both the 17th and 18th bands have been observed to comprise the major FS sheets although their topology is quite different from each other. For the magnetic borocarbides, the `cushion'-like FS from the 18th band, coming exclusively from the Ni 3$d_{x^2-y^2}$ and 3$d_{xy}$ derived states, is not affected by the magnetic moments of the rare earth ions and, thus, the superconductivity originating from this band can survive under the development of magnetic order.\cite{Drechsler-2004PhysicaC, BBergk-2008PRLdHvA}

The superconductivity in the non-magnetic borocarbides RNi$_2$B$_2$C (R = Y, Lu) has been associated with both the 17th and 18th bands.\cite{BBergk-2008PRLdHvA} Thus, a fundamental question to be addressed is comprised of two parts: (i) whether the superconducting order parameter consists of multiple components in these bands; and (ii) what is the order parameter symmetry and anisotropy in momentum space. Early on, Shulga {\it et al.}\cite {shulga-1998-80} reported on the evidence for a multi-band nature of the superconductivity from the temperature dependence measurements of the upper critical field in YNi$_2$B$_2$C and LuNi$_2$B$_2$C. A variety of experimental investigations such as photoemission spectroscopy\cite{Yokoya-2000-85}, Raman scattering \cite {Yang-2000-62}, thermal conductivity\cite {Boaknin-2001-87}, specific heat\cite{izawa01}, and ultrasound attenuation\cite{watanabe04} point to a large anisotropy in the superconducting gap functions. More detailed information on the gap anisotropy is reported from the measurements of thermal conductivity \cite {izawa02, YMatsuda-2006JPhy} and specific heat \cite {park-2003-90,park-2004-92} as a function of magnetic field orientation. Clear four-fold oscillations are observed in both kinds of measurements and interpreted as strong evidence for an anisotropic gap structure with gap minima located along the [100] and [010] directions. In particular, Izawa {\it et al.}\cite {izawa02, YMatsuda-2006JPhy} report the possible existence of point-like nodes from the observation of diminishing oscillation amplitude as the field direction changes from the $ab$-plane to the $c$-axis. This may be consistent with an extremely large gap anisotropy: $\gamma_\textrm{gap} \equiv \frac{\Delta_\textrm{max}}{\Delta_\textrm{min}}\geq$100. It is noted that these measurements can be obscured by the anisotropy in properties other than the superconducting order parameter, making their interpretations non-trivial.\cite{udagawa05,vorontsov06} This can be an issue particularly relevant to the borocabide superconductors since many properties are known to be anisotropic such as the FS,\cite{Dugdale-1999-83, BBergk-2008PRLdHvA, starowicz-1, SBDugdale-09SuperSciTech} upper critical field,\cite{metlushko97} and vortex lattice structure.\cite{dewilde97,eskildsen97,nakai02, nishimori04}

Maki and coworkers\cite{Maki-2002-65, QSYuan-2003PRB-1} propose a hybrid s+g wave order parameter, in which the g-wave component is added as an ansatz to simulate the point-node structure implied by experimental observations. Kontani\cite{Kontani-2004-70} point out that point-like nodes can be formed in a phonon-mediated s-wave superconducting order parameter if antiferromagnetic fluctuations exist on some parts of the FS connected by a nesting vector. Earlier, band structure calculations based on local density approximation predicted a nesting feature on the FS with nesting vectors $\textbf {Q}\approx 2\pi (0.5/a,0)$ and $2\pi(0,0.5/a)$ in the basal plane.\cite{JYRheee-95PRB-nesting} This is directly observed from two-dimensional angular correlation measurements on LuNi$_2$B$_2$C using the electron-positron annihilation radiation technique.\cite {Dugdale-1999-83} Interestingly, an antiferromagnetic order in the magnetic borocabide compounds is observed to have an ordering wave vector $\textbf{Q}_m\approx 2\pi (0.55/a,0,0)$,\cite{AIGoldman-PRB94, grigereit94, JZarestky-PRB95} very close to the nesting vector. The phonon-mediated pairing has long been attributed to the superconductivity in the borocarbide family.\cite{lawrie95,cheon99} For instance, Martinez-Samper {\it et al.}\cite{Martinez-Samper-2003-67} report on scanning tunneling spectroscopy measurements, in which features due to strong electron-phonon coupling reminiscent of a Pb tunnel junction are observed. Their extracted coupling constant is highly anisotropic due to soft phonon modes as observed in inelastic neutron scattering experiments.\cite{dervenagas95, bullock98, zarestky99} According to the theoretical arguments by Kontani,\cite{Kontani-2004-70} the nesting feature in the FS would generally weaken the electron-phonon coupling along $\textbf {Q}$. Thus, this scenario can explain the experimentally observed anisotropic gap structure in the nonmagnetic borocarbide superconductors. 

Spectroscopy based on the measurement of Andreev reflection conductance has been adopted frequently as a measure to investigate superconducting order parameters (see refs.~\onlinecite{park09} and~\onlinecite{park08} and references therein). Called point-contact Andreev reflection spectroscopy (PCARS), it is a simple and versatile technique, with which two groups have reported results on the nonmagnetic borocarbide family. Disparate claims are made from measurements using virtually the same technique. Raychaudhuri {\it et al.}\cite {Raychaudhuri-2004-93} observe a large anisotropy ($\gamma_\textrm{gap} \approx 5$) between the gap values along the [100] and [001] directions. They interpret this as evidence for the s+g order parameter with point nodes as suggested by Maki {\it et al.}\cite{Maki-2002-65} Later, they interpret this as evidence for multiband superconductivity.\cite {mukhopadhyay-2005-72} Naidyuk and coworkers do not report such large anisotropy.\cite{naidyuk-2007,naidyuk07,bobrov-2005-71,bashlakov05,bashlakov07} Between these two groups, the temperature dependences of the gap value along the [100] direction are strongly contradictory. Thus, our motivation for this work was, in part, to address this controversy with our own detailed gap structure measurement in the non-magnetic borocarbide superconductors.\cite{lu08}

\section{Experiments}

The working principle behind point-contact spectroscopy is that the energy dependence of quasiparticle scattering at an interface between two metallic electrodes is reflected as a non-linearity in the current-voltage characteristics. Analysis of such nonlinearity provides important information on the scattering sources. When applied to a superconductor, the most relevant scattering process is Andreev reflection, which is essentially a scattering off the pair potential. Thus, PCARS relies on the strong energy dependence of this scattering at a normal-metal/superconductor interface. In order to avoid complications due to other scattering processes (elastic or inelastic), a point-contact junction must fulfill requirements for the ballisticity. Ideally, its size needs to be smaller than the electronic mean free paths of the electrodes. In practice, nanoscale metallic junctions can be made by bringing a sharpened metal tip into contact with a superconductor using fine mechanical adjustments.

We perform systematic investigations of the gap structures in the nonmagnetic borocabide superconductors YNi$_2$B$_2$C and LuNi$_2$B$_2$C. Differential conductance spectra along three major crystallographic directions are taken from two sets of LuNi$_2$B$_2$C single crystals of different sources and one set of YNi$_2$B$_2$C single crystals. Crystals from different batches are used for each direction in the set LuNi$_2$B$_2$C \#1. For the sets LuNi$_2$B$_2$C \#2 and YNi$_2$B$_2$C, crystals for all three directions are prepared from same batches. These single crystals are grown by flux method using Ni$_2$B flux.\cite{xu94, canfield01} Bulk resistivity and magnetization measurements show that they are of high quality, as summarized in Table I.

\begin{table*}[htbp]
\caption{\label{table:table1} Summary of sample characteristics and PCARS results. RRR (residual resistance ratio) $\equiv R(300\textrm{K})/R(T_\textrm{c,on})$, ($T_\textrm{c,on}$ = onset transition temperature); $T_\textrm{c,zero}$: zero-resistance transition temperature; $\Delta T_\textrm{c} \equiv T_\textrm{c,on} - T_\textrm{c,zero}$; $\Delta_0$: gap energy at zero-temperature; $\gamma_\textrm{gap} \equiv \Delta_{\textrm{max}}/\Delta_{\textrm{min}}$: maximum gap anisotropy; $\omega \equiv 1-\Delta_{001}/\Delta_{110}$: weight for g-wave component (see text).}
\vspace{5pt}
\begin{tabular}{p{0.9in}p{0.7in}p{0.6in}p{0.6in}p{0.6in}p{0.6in}p{0.6in}p{0.6in}p{0.6in}} 
\hline\hline
Crystal & Orientation & RRR &  $T_\textrm{c,zero}$(K) & $\Delta T_\textrm{c}$(K)  & $\Delta_0$(\textrm{meV}) & 2$\Delta_0/k_\textrm{B}T_\textrm{c}$ & $\gamma_\textrm{gap}$ & $\omega$ \\
\hline
                    & [001] & 21   & 16.0  & 0.7   & 2.4  & 3.48  &      &       \\[0.2ex]
{LuNi$_2$B$_2$C \#1}& [110] & 21   & 16.1  & 0.6   & 2.6  & 3.75  & 1.13 & 0.077 \\[0.2ex]
                    & [100] & 20   & 15.0  & 0.5   & 2.3  & 3.56  &      &       \\[1ex]

                    & [001] &      &       &       & 2.5  & 3.60  &      &       \\[0.2ex]
{LuNi$_2$B$_2$C \#2}& [110] & 26   & 16.1  & 0.7   & 2.8  & 4.04  & 1.12 & 0.107 \\[0.2ex]
                    & [100] &      &       &       & 2.7  & 3.89  &      &       \\[1ex]

                    & [001] &      &       &       & 2.0  & 3.00  &      &       \\[0.2ex]
{YNi$_2$B$_2$C}     & [110] & 24   & 15.5  & 0.3   & 2.1  & 3.15  & 1.25 & 0.048 \\[0.2ex]
                    & [100] &      &       &       & 2.5  & 4.04  &      &       \\[0.2ex]
\hline
\end{tabular}

\label{tab:PPer}
\end{table*}

For a point-contact junction oriented along the $c$-axis, the as-grown surface of a single crystal is used, since its normal is along that direction. For in-plane junctions along [100] and [110] directions, crystals are embedded into low-temperature epoxy, cut, and polished until atomic smoothness (rms roughness $\sim$10 \AA) is achieved.\cite{lu08,park07,park08} The desired crystallographic orientations of the exposed surfaces are identified by X-ray diffraction, with 5 degree accuracy with the intended directions obtained. Since pristine crystals may contain a degraded surface layer where superconductivity is strongly suppressed, which is not favorable to PCARS due to its surface-sensitive nature, the crystals are etched slightly in aqua-regia for about 20 seconds to expose fresh surfaces prior to making point-contact junctions. For the PCARS measurement, a crystal is cooled down to about 2 K in the liquid helium cryostat, and then a sharp gold tip is moved to engage onto the crystal surface by a fine differential micrometer until a reasonable junction resistance is reached.

The junction resistance, $R_\textrm{J}$, usually ranges from several ohms to tens of ohms. Applying the Wexler's formula\cite{wexler66} using these $R_\textrm{J}$ values along with known materials parameters for borocarbides ($\rho_0\sim$ 1.30 $\mu\Omega \cdot$cm and $l\sim$ 500 \AA \ as reported in ref.~\onlinecite{Boaknin-2001-87}), we infer that the junctions are in the ballistic limit. What is more important to ensure the spectroscopic nature is that our measured conductance spectra are reproducible as shown below and free from non-spectroscopic effects such as local heating. The differential conductance ($G \equiv dI/dV$) as a function of voltage ($V$) is directly recorded by the standard four-probe lock-in technique over wide temperature ($1.6 - T_\textrm{c}$) and magnetic field ($0 - 9$ T) ranges covering the whole phase space for the superconductivity in these materials. The local critical temperature in the junction area, $T_\textrm{c}^j$, is determined by the temperature dependence of the zero-bias conductance. The measured $T_\textrm{c}^j$ is in good agreement with the bulk $T_\textrm{c}$ determined by four-probe resistance measurement. This is an indication that the properties of the bulk rather than a possible degraded surface layer are measured in our PCARS experiments. For brevity, we simply refer to $T_\textrm{c}^j$ as $T_\textrm{c}$ in the following.

\section{Superconducting Energy Gap}

\subsection{LuNi$_2$B$_2$C}

\begin{figure*}[tp]
\includegraphics[angle=0,width=0.95\textwidth]{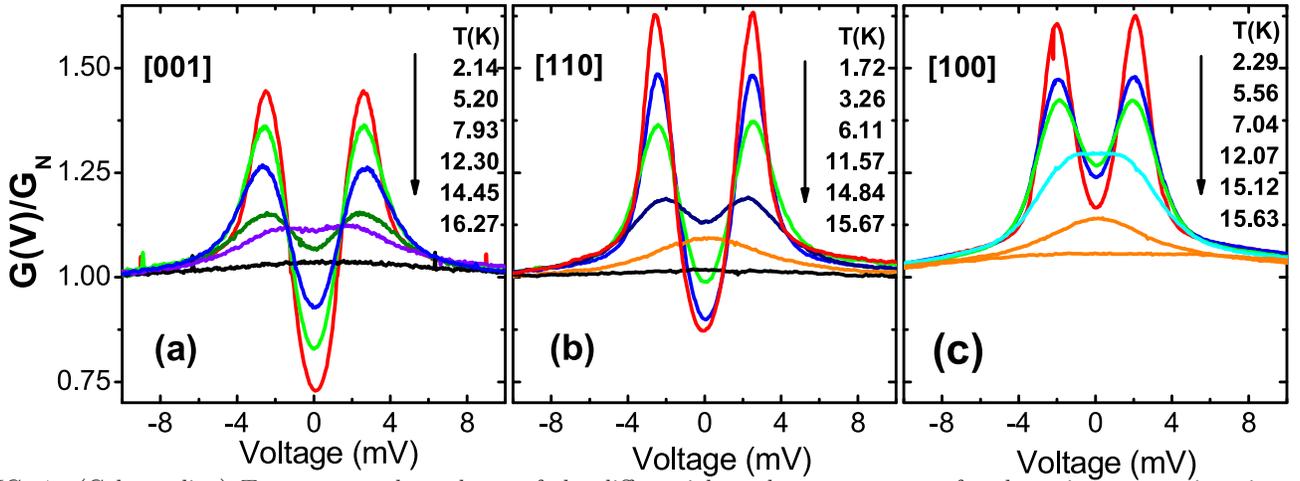}
\vspace{-15pt}
 \caption{\label{fig:fig1} (Color online) Temperature dependence of the differential conductance spectra for the point-contact junctions on LuNi$_2$B$_2$C \#1 along (a) [001], (b) [110] and (c) [100] directions. These $G(V)$ curves are obtained from junctions showing the most frequent gap values in Fig. 3. The data are normalized by the conductance at the negative maximum bias voltage in (a), (b), and (c).}
\end{figure*}

Figure \ref{fig:fig1} shows one set of $G(V)$ curves for LuNi$_2$B$_2$C \#1 crystals, taken as a function of temperature along the three major crystallographic orientations: [001], [110], and [100]. Note that at the lowest temperatures the Andreev reflection amplitude and peak positions in the three directions are comparable. The data are analyzed using the Blonder-Tinkham-Klapwijk (BTK) model\cite{blonder-1982-25,Plecenik-1994-49} assuming a single isotropic s-wave gap. In this BTK model, there are three parameters: the superconducting gap ($\Delta$) extracted from the BTK fit, the barrier strength ($Z$) between the normal metal and superconductor, and the quasiparticle scattering parameter ($\Gamma$), which accounts for the smearing of a conductance curve due to shortened quasiparticle lifetime or depairing effect caused by inelastic scattering or magnetic field. Here, we focus on the extracted superconducting gap and its temperature dependence.

\begin{figure}[tp]
\begin {center}
\includegraphics[angle=0,width=0.48\textwidth]{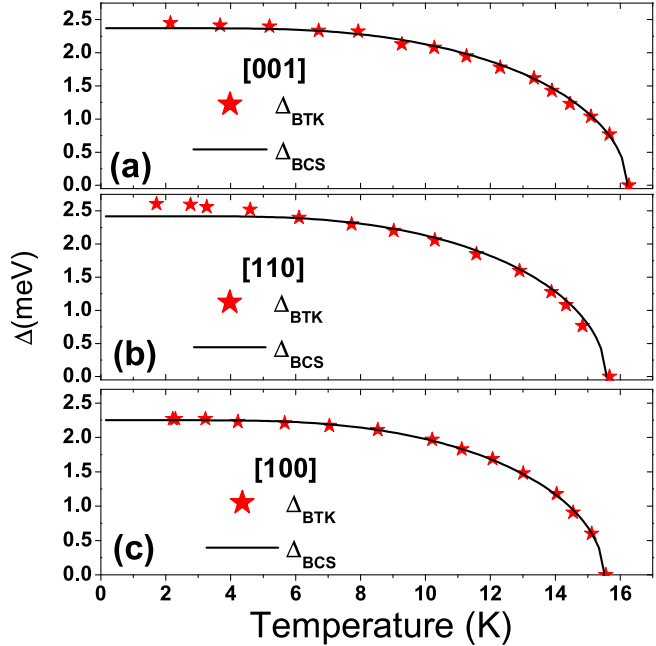}
\end {center}
\vspace{-15pt}
\caption{\label{fig:fig2} (Color online) Temperature dependence of the superconducting gap in LuNi$_2$B$_2$C \#1 extracted from an analysis using the single-gap BTK model along (a) [001], (b) [110] and (c) [100] directions. The symbols are experimental data and the solid lines represent the BCS prediction.}
\end{figure}

\begin{figure}[tp]
\begin {center}
\includegraphics[angle=0,width=0.48\textwidth]{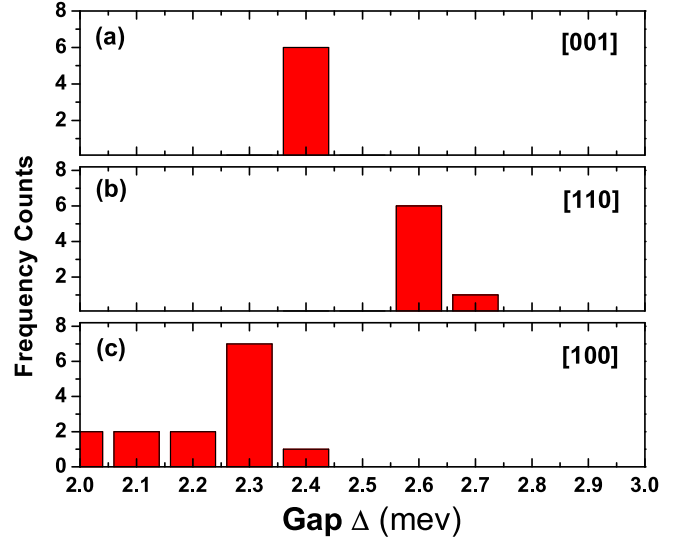}
\end {center}
\vspace{-15pt}
\caption{\label{fig:fig3} (Color online) Histograms plotting the frequencies of the superconducting gap values measured from point-contact junctions on LuNi$_2$B$_2$C \#1 along (a) [001], (b) [110], and (c) [100] direction.}
\end{figure}

Figure \ref{fig:fig2} shows the temperature dependence of the superconducting gap, $\Delta$, in the [001], [110], and [100] direction, respectively. As shown, they all follow the standard BCS-like curve, yielding $2\Delta_0/k_BT_\textrm{c}\sim$ 3.5, 3.8 and 3.6 in the weak-coupling limit. This is in good agreement with the results obtained from break junction experiments.\cite{ekino96} At the lowest measurement temperatures ($\sim$ 2 K), the [001] and [110] junctions show comparable superconducting gap values with $\Delta_{001}=2.4$ meV and $\Delta_{110}=2.6$ meV, respectively. The extracted gap value along the [100] direction is $\sim$ 2.3 meV, giving a small anisotropy with the other directions ($\gamma_\textrm{gap}\sim$1.13). As shown in Table I, the bulk $T_\textrm{c}$ of this crystal is lower than those for the others. Its origin is not clear yet but the 2$\Delta_0/k_\textrm{B}T_\textrm{c}$ value is consistent within the LuNi$_2$B$_2$C \#1 set. Our gap values are similar to those reported by Bobrov {\it et al.},\cite {bobrov-2005-71} where the $c$-axis and $ab$-plane gap sizes from one-gap BTK fit are 2.25 and 2.55 meV, respectively. We note that their samples for $ab$-plane junctions are prepared without well-defined orientations. In our study, dozens of point-contact junctions have been measured and their conductance spectra are quite reproducible as demonstrated by the histograms in Fig. \ref{fig:fig3}, which counts the occurrences for different gap values at the lowest measurement temperatures. $\Delta_{100}$ exhibits a bit scattered distribution although the most frequently observed value is 2.3 meV. This behavior might be indicative of the tunneling cone effect, which is most pronounced along the minimum gap direction, as discussed later.  
   
The superconducting energy gaps obtained from a single-band BTK analysis show slight deviations from the BCS curve in the low temperature region, as seen in Fig. 2(b) and in other data in the following. We also note that in the low temperature region the best-fit curves are not as satisfactory as for a known isotropic single-band superconductor such as Nb. Bobrov \textit{et al.}\cite {bobrov-2005-71} have reported similar observations in their PCARS on LuNi$_2$B$_2$C crystals and demonstrated that their $G(V)$ curves can be better fit by assuming a continuous distribution of the gap function with double maxima. They interpreted this as evidence for an anisotropic gap structure or multiband superconductivity in LuNi$_2$B$_2$C. Although such an analysis is not an unreasonable approach considering recent developments in the field, we view it as marginal since it does not provide corroborative evidence for multiple order parameters. Note that strong evidence for multiband superconductivity in MgB$_2$ does not come from improved conductance fitting but from clear features due to multiple gaps as observed in the original data.\cite{szabo01,park06}    

\begin{figure}[tp]
\includegraphics[angle=0,width=0.48\textwidth]{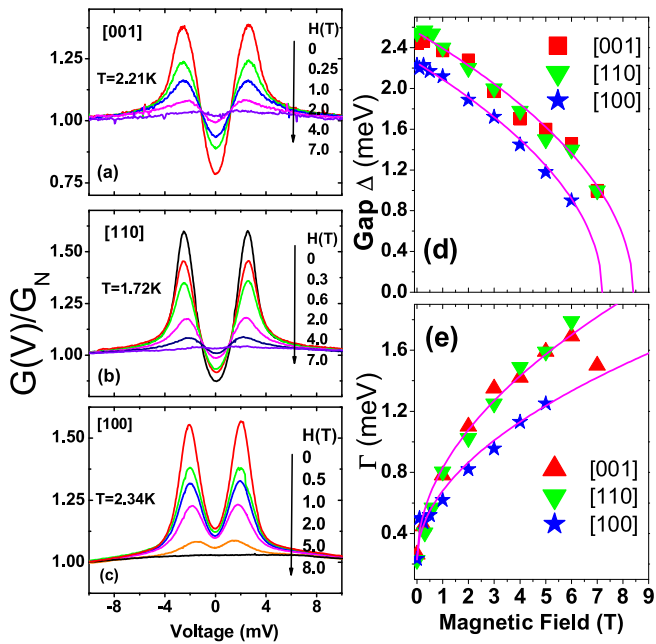}
\caption{\label{Lu1221-Fielddep} (Color online) Magnetic field dependence of the differential conductance for the point-contact junctions on LuNi$_2$B$_2$C \#1 along (a) [001], (b) [110] and (c) [100] directions. The data are normalized by the conductance at the negative maximum bias voltage in (a), (b), and (c). The gap $\Delta$ extracted from a single-gap BTK analysis is plotted in (d) and the quasiparticle broadening factor $\Gamma$ in (e), as a function of the magnetic field. The solid lines in (d) and (e) represent best-fit curves using $\Delta = \Delta_0 (1-H/H_\textrm{c2})^{1/2}$ and $\Gamma=\Gamma_0+c_0\sqrt{H}$, respectively. Here, $\Gamma_0 = 0.23$ meV for all three junctions; $c_0 = 0.60$ meV $\cdot$ T$^{-1/2}$ for the [001] and [110] junctions, and $c_0 =$ 0.45 meV $\cdot$ T$^{-1/2}$ for the [100] junction.}
\end{figure}

The magnetic field dependence of the superconducting gap was studied at the lowest temperatures around 2 K, as shown in Fig. \ref{Lu1221-Fielddep}. The $G(V)$ curves were taken with the field applied along the corresponding crystallographic direction. With increasing field, the conductance peak is suppressed and broadened with its position moving toward the zero bias. As shown in Fig.\ref{Lu1221-Fielddep} (d), the superconducting gap decreases with increasing field, roughly following the BCS behavior, $\Delta = \Delta_0 (1-H/H_\textrm{c2})^{1/2}$. No field-dependent signatures for multiple gaps are observed. The $\Gamma$ parameter increases with $\sqrt{H}$ dependence, as plotted in Fig.\ref{Lu1221-Fielddep} (e).  

\begin{figure}[tp]
\includegraphics[angle=0,width=0.48\textwidth]{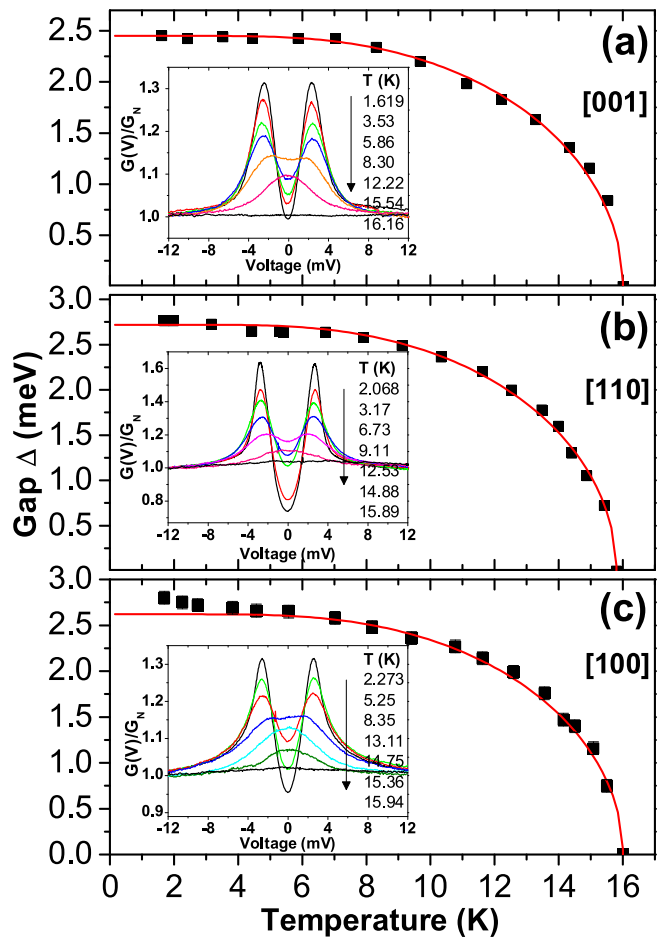}
\vspace{-20pt}
\caption{\label{fig:fig44} (Color online) Temperature dependence of the superconducting gap extracted from a single-gap BTK analysis for point-contact junctions on LuNi$_2$B$_2$C \#2 along (a) [001], (b) [110] and (c) [100] directions in comparison with the standard BCS curve. The insets plot the conductance data as a function of temperature.}
\end{figure}

Another set of LuNi$_2$B$_2$C crystals from a different source are prepared in the same way in order to check sample dependence. The $G(V)$ curves and the extracted gap values are displayed in Fig. \ref{fig:fig44} as a function of temperature. Similar behaviors in the gap function to the previous data set are observed, following the BCS prediction and showing a small anisotropy. We note that $2\Delta_0/k_BT_\textrm{c}$ values for these samples are systematically larger than those in the first set by $\sim (3-9)\%$ although their $T_\textrm{c}$ values are comparable, as listed in Table. This means that they are in the intermediate coupling regime. What causes this difference between the two sets of LuNi$_2$B$_2$C crystals is not clear. 

\subsection{YNi$_2$B$_2$C}

In terms of electronic structure, there is not much difference between YNi$_2$B$_2$C and LuNi$_2$B$_2$C.\cite{Yamauchi-2004-412, BBergk-2008PRLdHvA} Although the maximum bulk $T_\textrm{c}$ differs by 1 K, this can be accounted for by a small difference in the electron-phonon coupling constant or the density of states at the Fermi level. The superconducting gap has been reported to be anisotropic in both compounds. Nonetheless, we have carried out PCARS measurements on YNi$_2$B$_2$C single crystals in order to rule out the possibility that the gap anisotropy may be somehow enhanced in this compound than in LuNi$_2$B$_2$C. As shown in Table \ref{table:table1}, the RRR values of these crystals are comparable to those of LuNi$_2$B$_2$C crystals. 
 
\begin{figure}[tp]
\includegraphics[angle=0,width=0.48\textwidth]{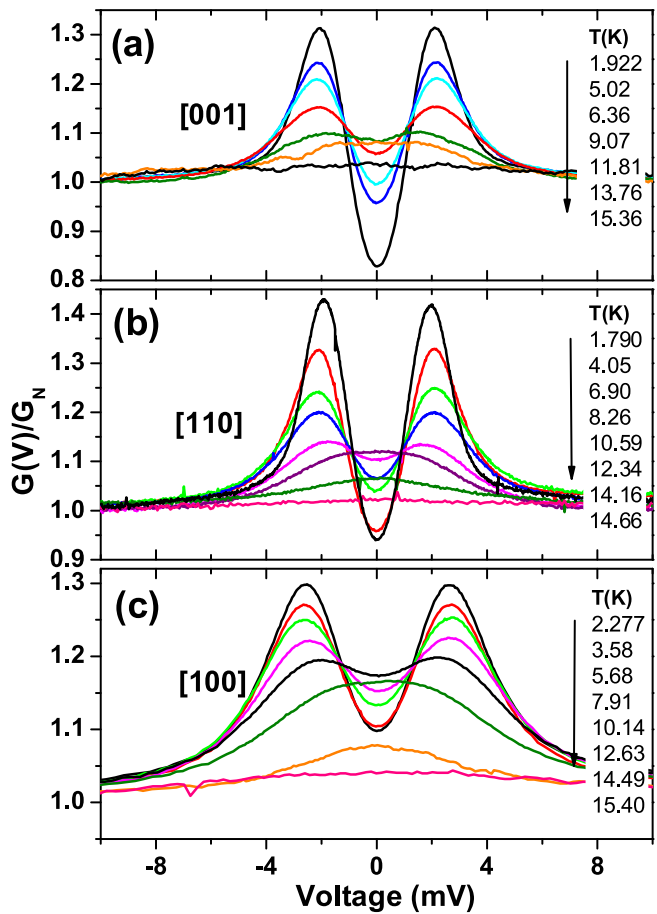} 
\vspace{-15pt}
\caption{\label{4-3-2_PCSTemp} (Color online) Temperature dependence of the normalized differential conductance for the point-contact junctions on YNi$_2$B$_2$C along (a) [001], (b) [110] and (c) [100] directions. The data are normalized with respect to the negative maximum bias voltage.}
\end{figure}
 
\begin{figure}[tp]
\includegraphics[angle=0,width=0.48\textwidth]{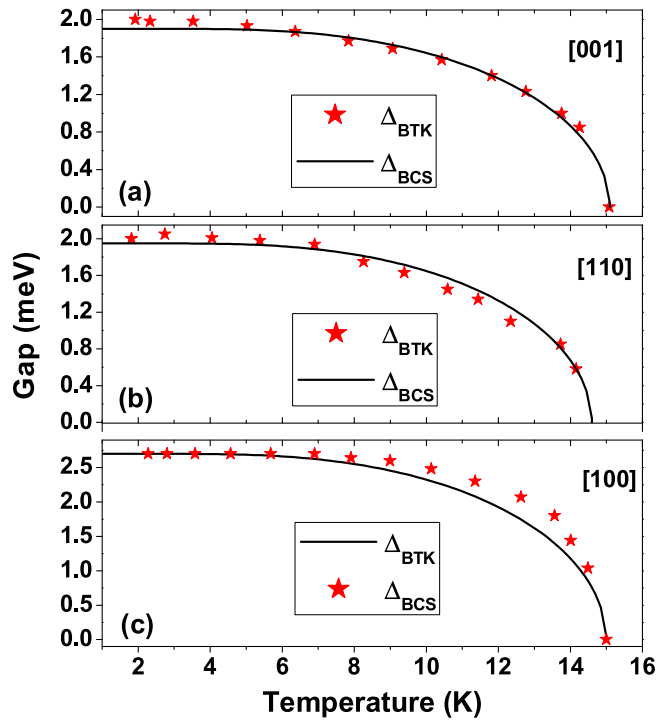}
\vspace{-15pt}
\caption{\label{4-3-3_GapTemp} (Color online) Temperature dependence of the superconducting gap in YNi$_2$B$_2$C along (a) [001], (b) [110] and (c) [100] directions in comparison with the standard BCS curve.}
\end{figure}

The temperature evolution of the conductance spectra and the gap energy from point-contact junctions on YNi$_2$B$_2$C are shown in Fig. \ref{4-3-2_PCSTemp} and Fig. \ref{4-3-3_GapTemp}, respectively. Again, these data are analyzed with the single-gap BTK model to extract the gap values. They all follow the BCS curve although some deviations are observed in the low temperature region, similarly to those observed in LuNi$_2$B$_2$C. The averaged gap values obtained from dozens of contacts at the lowest measurement temperatures of $\sim$ 2 K are given in Table \ref{table:table1}. The maximum gap anisotropy $\gamma_\textrm{gap}$ is 1.25, larger than in LuNi$_2$B$_2$C, but the gap is maximum along the [100] direction. This is in contrast to the case of LuNi$_2$B$_2$C, where the gap is maximum along the [110] direction. 

As in the case of LuNi$_2$B$_2$C, the extracted energy gap closes at the bulk resistive $T_\textrm{c}$ in all three directions. This is a strong indication that our PCARS results represent bulk properties instead of a possible surface layer that is degraded during experiments. Theoretically, it is possible that the small gap anisotropy may be caused by the averaging effect due to microscopic surface facets. In order to check for this possibility, we took conductance data on crystals with and without undergoing the chemical etching process after polishing. Similar results are obtained from both cases, ruling out such possibility. 
        
\section{Discussions}

As presented above and summarized in Table \ref{table:table1}, we have observed a small gap anisotropy from both nonmagnetic superconductors LuNi$_2$B$_2$C and YNi$_2$B$_2$C. Maximum gap anisotropy ranges between 1.12 and 1.25. This observation is reproducible and robust among different sets of samples and measurements. Our results are apparently contradictory with many recent reports in the literature claiming for a strong anisotropy in the gap structure, as discussed in Sec. I.\cite{Yokoya-2000-85, Yang-2000-62, Boaknin-2001-87, izawa01, watanabe04, izawa02, YMatsuda-2006JPhy, park-2003-90, park-2004-92} Below, we put forward possible explanations for this discrepancy, based on the Fermi surface topology and tunneling cone effect. We note that, by nature, thermal conductivity and specific heat in the superconducting state are sensitive to the gap minimum, whereas inherently anisotropic gap features could be smeared out in PCARS since the momentum of an injected electron can distribute over a wide range of angle with respect to the nominal surface normal. We begin by reviewing the Fermi surface topology revealed by band structure calculations and recent angle-resolved photoemission spectroscopy (ARPES),\cite{TBaba-ARPES} and quantum oscillation experiments.\cite{BBergk-2008PRLdHvA}

The claimed nesting structure on the 17th band FS has been observed by different experiments such as ARPES,\cite{starowicz-1} and electron-positron annihilation measurements \cite{SBDugdale-09SuperSciTech}, where the nested part only occupies a small portion of the Fermi surface.\cite{Drechsler-2004PhysicaC} According to Kontani,\cite{Kontani-2004-70} antiferromagnetic fluctuations on the nested parts of the FS could induce a point-node-like gap minimum along the nesting wave vector. Recently, Baba {\it et al.}\cite{TBaba-ARPES} have reported on the detailed gap structures in YNi$_2$B$_2$C measured by high-resolution ARPES. Their results show that the superconducting gap is observed on multiple bands (the 17th and 18th) with very different momentum dependences. More specifically, it is revealed that the gap is highly anisotropic on the 17th band with two minima, whereas the 18th band has a nearly constant gap. The two minima are 1.5 and 2.3 meV on different parts of the 17th band FS. Their momentum directions are identified as [100], consistent with other reports.\cite{izawa02, YMatsuda-2006JPhy, park-2003-90, park-2004-92}
It was claimed that the points on the FS where the gap shows a minimum of 1.5 meV can be connected by the nesting vector $\textbf{Q} \sim 2\pi/a (0.55, 0, 0)$, implying an intimate connection between the minimal gap and the FS nesting. It was also argued that the reason why the gap minimum has a non-zero value is because their ARPES data were taken off the  basal plane with $k_z\sim 0.5 \frac{2\pi}{c}$. 

Based on these observations from ARPES, it is naturally understood that the node-like gap minimum along [100] cannot be observed in PCARS since the shape of a conductance curve will be dominated by the other gap minimum (2.3 meV) which is located along the same direction. Indeed, this gap value is close to those we observed in this study (see Table \ref{table:table1}.), explaining why we observe only a small gap anisotropy in both LuNi$_2$B$_2$C and YNi$_2$B$_2$C.

Our experimental results are in strong contradiction with those reported by Raychaudhuri and coworkers.\cite{Raychaudhuri-2004-93, mukhopadhyay-2005-72} From PCARS measurements on YNi$_2$B$_2$C, they reported a gap anisotropy of $\gamma_\textrm{gap} \sim 4.5$, much larger than ours ($1.12 - 1.25$). This discrepancy is mainly associated with their extremely small gap value along [100] ($\sim 0.415$ meV) since their gap value along [001] ($\sim 1.8$ meV) is not much different from ours. Moreover, they observed that $\Delta_{100}$ closes at $\sim (7 - 9)$ K, much lower than the bulk $T_\textrm{c}$. This is in strong contrast with our observation of $\Delta_{100}$ closing at the bulk $T_\textrm{c}$. We note that Naidyuk {\it et al.}\cite{naidyuk-2007} also reported similar temperature dependences to ours. Raychaudhuri {\it et al.} originally interpreted their results as evidence for s+g gap structure with point nodes,\cite{Raychaudhuri-2004-93} and later on claimed for the multiband nature of the superconductivity in borocarbides.\cite{mukhopadhyay-2005-72} We think this is an unlikely interpretation considering the recent ARPES measurements discussed above since there is no way for PCARS, which doesn't have such high momentum resolution as in ARPES, to probe only an  extremely small gap value ($\sim 0.415$ meV) while there exists another but sizable gap minimum ($\sim 2.3$ meV) along [100]. Rather, it is more natural to attribute the exotic behavior of their measured $\Delta_{100}$ to non-intrinsic effects such as suppressed superconductivity on the sample surface, as was also pointed out by Naidyuk {\it et al.}\cite{naidyuk-2007}

Next, we discuss the tunneling cone effect on the conductance spectra. In the original BTK model,\cite{blonder-1982-25} the integration is carried out only over the energy, assuming an isotropic gap structure. Thus, it is inaccurate for the analysis if the energy gap has substantial dependence on momentum. A straightforward way to deal with this situation is to include an integration with respect to momentum. Here sets in the tunneling cone effect.\cite{beuermann81,wolf85} If we consider a planar tunnel junction with a potential barrier of finite thickness, the transmission probability varies depending on the momentum direction, being maximal for a direction normal to the barrier. Assuming a Gaussian distribution for the probability, the tunneling cone angle can be as small as $5 - 10$ degrees in typical tunnel junctions.\cite{wolf85} This angle is expected to increase with decreasing barrier strength. Therefore, its effects are not negligible in PCARS due to inherently small barrier strength as well as small junction size (uncertainty principle). Here, we simulate conductance spectra in order to study the effect of tunneling cone on the gap structure with large anisotropy. Namely, we calculate conductance curves that are obtained from a given gap function with a varying cone angle, and then determine the gap values from the peak positions.

\begin{figure}[tp]
\includegraphics[angle=0,width=0.48\textwidth]{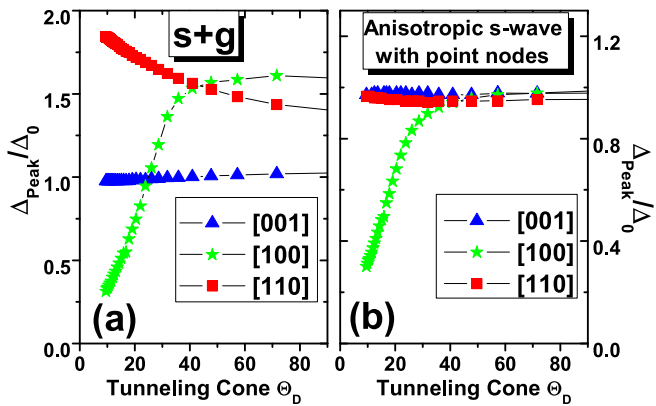}
\vspace{-15pt}
\caption{\label{fig:tcone} (Color online) Dependence of the peak position, $\Delta_{Peak}/\Delta_0$, of the conductance curve simulated by the extended BTK model on the tunneling cone angle, $\Theta_D$, assuming a nodal gap structure of (a) s+g wave and (b) anisotropic s-wave, respectively.}
\end{figure}

In order to simulate conductance curves, we adopt the extended BTK model for a $d$-wave superconductor, formulated by Tanaka and Kashiwaya,\cite {kashiwaya-1996-53} but only consider gap functions without a sign change since there is substantial evidence in these materials for this premise. Then, the conductance kernel $\sigma_S(E)$ for a given $\theta_N$, angle of incidence in the normal-metal electrode, can be written as:
\begin{equation}
\sigma_S(E)=\sigma_N \frac{1+\sigma_N|\Gamma_+|^2+(\sigma_N-1)|\Gamma_+\Gamma_-|^2}{|1+(\sigma_N-1)\Gamma_+\Gamma_-|^2},
\label{eq:kernel}
\end{equation}
where $\sigma_N=\frac{1}{1+Z^2}$, $Z=\frac{Z_0}{\cos \theta_N}$ and
$\Gamma_{\pm}=\frac{E-\sqrt{E^2-|\Delta_{\pm}|^2}}{|\Delta_{\pm}|}$. Here $Z_0$ is the barrier strength, and $\Delta_{\pm}$ represent the pairing potentials for the transmitted electron-like quasiparticles or hole-like quasiparticles, respectively. Thus, for a normal-metal/superconductor junction, the total conductance $\sigma_T(E)$ is given by the integration of $\sigma_S(E)$ with respect to a solid angle $\Omega$,
\begin{equation}
\sigma_T(E)=\frac{\int d\Omega \sigma_S (E) \cos \theta_N P(\theta_N)}{\int d\Omega \sigma_N \cos \theta_N P(\theta_N)},
\label{eq:solid angle}
\end{equation}
where $P(\theta_N)$ is the transmission probability for a given $\theta_N$. As an approximation, we assume a Gaussian-type distribution:
\begin{equation}
P(\theta_N) \varpropto e^{-(\frac{\theta_N}{\Theta_D})^2}.
\label{eq:tcone}
\end{equation}
The parameter $\Theta_D$ is a variable characterizing the tunneling cone for a given junction.
We consider two gap functions: s+g wave as proposed by Maki {\it et al.}\cite{Maki-2002-65} and anisotropic s-wave gap, both having four-fold point-nodes. The s+g gap function can be written as $\Delta({\bf k})=\frac{1}{2} \Delta_0 [1-\sin^4\theta \cos(4\phi)]$ and for an anisotropic gap we assume $\Delta({\bf k})=\Delta_0[1-\sin^4 \theta \cos^2(2\phi)]$ for simplicity. In the zero-temperature limit, the peak position of a conductance curve can be taken as the energy gap. Figure \ref{fig:tcone} plots thus obtained $\Delta$ along three major directions as a function of the tunneling cone angle, $\Theta_D$. An assumption is made that the tunneling cone $\Theta_D$ does not depend on the surface orientations. The inherently large anisotropy is observed for a narrow tunneling cone. However, the anisotropy becomes smaller as the cone angle increases. $\Delta_{100} = \Delta_{001}$ at $\Theta_D$ =25 and 35 deg., for the s+g and anisotropic s-wave case, respectively. Although qualitatively similar, there is difference between the two gap functions in their response to the tunneling cone. That is, the calculated gap values do not merge in the s+g case, whereas they merge into one for large tunneling cone in the anisotropic s-wave case. This indicates that the anisotropic s-wave with $\Theta_D \geq 30$ deg., rather than the s+g, agrees with the small gap anisotropy we have observed constantly. If we consider more general case of the s+g gap function, written as $\Delta = \Delta_0 [(1-\omega)-\omega \sin ^4\theta \cos(4\phi)]$, we can estimate the weight for the g-wave component. It turns out that $\omega$ is very small in all our measurements as listed in Table \ref{table:table1}. Thus, we argue that the s+g wave is unlikely as a candidate gap function in borocarbides, in contrast to several reports in the literature.\cite{izawa02,YMatsuda-2006JPhy, Raychaudhuri-2004-93}

Our simulation based on tunneling cone effect is essentially similar to the BTK fitting with a distribution of gap values as reported by Bobrov {\it et al.}\cite{bobrov-2005-71} in that contributions from different gap values are added in both cases. As we do not claim for an anisotropic gap structure in the borocarbide superconductors from our tunneling cone simulation, we don't view their arguments as corroborative evidence for the multiband superconductivity, either.

Studies on the thermal conductivity\cite{Kamata-2003} and specific heat\cite{nohara99} in doped single crystals of Y(Ni$_{1-x}$Pt$_x$)$_2$B$_2$C have shown that the gap anisotropy is washed out as the doping level increases. This was interpreted as due to gap opening induced by the nonmagnetic impurity, in agreement with theoretical calculations.\cite {QSYuan-2003PRB, Kamata-2003} Our transport measurements indicate that the samples used in our PCARS are of high quality and no correlation between RRR and $\gamma_{gap}$ is observed, as summarized in Table \ref{table:table1}. We note that the $\Delta_{100}$ value of a LuNi$_2$B$_2$C \#2 single crystal changes very little after annealing although its RRR increases to 34.3 (not shown here). Moreover, the specific heat results reported in refs.~\onlinecite{park-2003-90} and ~\onlinecite{park-2004-92} which were obtained using single crystals from the same sources as ours, have shown clear four-fold oscillations under magnetic field. Thus, the possibility of impurity scattering being an intrinsic origin for the small gap anisotropy can be excluded in our PCARS study.

Considering our experimental observations in relation to the discussions presented above, we predict that the anisotropic gap structure can be better probed by thin-film based tunneling spectroscopy. In particular, the two gap minima along the [100] direction as seen in the recent APRES study\cite{TBaba-ARPES} should also be discernible in the tunneling conductance data.

\section{Conclusions}     

We have carried out PCARS measurements on the nonmagnetic borocarbide superconductors LuNi$_2$B$_2$C and YNi$_2$B$_2$C in order to investigate their detailed gap structures. Conductance spectra as a function of crystallographic orientation, temperature, and magnetic field are taken from two different sets of LuNi$_2$B$_2$C crystals and one set of YNi$_2$B$_2$C crystals. Analysis based on the single-gap BTK model shows that the superconducting gap follows the BCS predictions as a function of temperature and magnetic field, closing at bulk superconducting transitions, albeit a slight deviation is sometimes observed. The measured superconducting gaps exhibit a small anisotropy among the three major directions: [001], [110], and [100]. Any clear evidence for multiple gaps is not observed. These results are completely reproducible and robust in all our measurements. In order to explain our observations, two possible scenarios are considered in terms of the recent ARPES studies and tunneling cone effect. We argue that these two effects may render it impossible to observe the inherently large gap anisotropy using PCARS. Thin-film tunneling spectroscopy on the [100] surface should better detect the gap minima, which then can serve as a confirmation of the anisotropic structure and multiband nature of the superconducting order parameter in these compounds. Our study addresses the controversy over the gap structure in the nonmagnetic borocarbides, particularly among the studies using PCARS techniques.

\begin{acknowledgments}

We acknowledge experimental help by undergraduate research assistants, A. S. Ahmed, V. Jacome, and Y.-H. Lin. The work at the University of Illinois at Urbana-Champaign was supported by the National Science Foundation under Award No. DMR 07-06013 (X.L.) and by the Department of Energy under Award No. DE-FG02-07ER46453 (W.K.P.) through the Frederick Seitz Materials Research Laboratory and the Center for Microanalysis of Materials. The work at the Ames Laboratory was supported by the Department of Energy, Basic Energy Sciences under Contract No. DE-AC02-07CH11358.
\end{acknowledgments}



\end{document}